\def\ROSAT{{\it ROSAT\/\ }}

\def\ASCA{{\it ASCA\/\ }}

\def\Ir{IRAS~13349+2438\ }
\def\Irc{IRAS~13349+2438}

\def\ltsima{$\; \buildrel < \over \sim \;$}
\def\simlt{\lower.5ex\hbox{\ltsima}}
\def\gtsima{$\; \buildrel > \over \sim \;$}
\def\simgt{\lower.5ex\hbox{\gtsima}}

\documentstyle[psfig]{mn}

\title[Hard spectral slopes of broad and narrow-line Seyfert 1 galaxies]
{A comparison of the hard \ASCA spectral slopes of broad and narrow-line Seyfert 1 galaxies}

\author[W.N. Brandt, S. Mathur \& M. Elvis]
{\parbox[]{6.5in}{W.N. Brandt, S. Mathur and M. Elvis}\\
\\
Harvard-Smithsonian Center for Astrophysics, 60 Garden Street, 
Cambridge, Massachusetts 02138, USA\\
}

\begin{document}

\maketitle

\begin{abstract}  
The soft ($\approx$ 0.1--2.0 keV) X-ray spectra of narrow-line 
Seyfert~1 galaxies are known to be generally steeper than those
of Seyfert~1 galaxies with broader optical permitted lines. This
has been attributed to the presence of strong soft X-ray
excesses, over the hard X-ray power law, in 
many narrow-line Seyfert 1s. Here we 
use the currently available \ASCA 
data to systematically compare the harder
($\approx$~2--10 keV) X-ray continua of soft \ROSAT narrow-line
Seyfert~1s with those of Seyfert~1s with larger H$\beta$ 
FWHM. Our robust and nonparametric testing suggests, 
with high statistical significance, that soft \ROSAT 
narrow-line Seyfert~1s have generally steeper 
intrinsic hard X-ray continua than Seyfert~1s with larger 
H$\beta$ FWHM. The hard photon index trend appears similar to 
the previously known soft photon index trend, although
with a reduced photon index spread. If the soft X-ray excesses 
of all Seyfert~1s are confined to below $\approx 1$~keV they
cannot directly affect the $\approx$~2--10~keV spectra
studied here. However, as suggested for the extreme narrow-line 
Seyfert~1 RE~J~1034+393, a strong soft X-ray excess may affect the
accretion disc corona which creates the underlying 
hard X-ray power law. If this is
occurring, then more detailed study of this physical process 
could give clues about the formation of the underlying continua 
of all Seyferts. Other effects, such as 
intrinsic 2--10~keV continuum 
curvature, could also lead to the observed photon
index trend and need further study.
\end{abstract}


\section{Introduction} 

The X-ray spectra of most Seyfert 1 type galaxies are
formed predominantly within $\sim 50$ Schwarzschild radii
of their black holes, while Seyfert optical permitted lines
are formed in a separate and significantly 
larger region. Thus, it was remarkable
when an extremely strong anticorrelation was found between \ROSAT
spectral softness and H$\beta$ full width at half-maximum
intensity (FWHM) in type 1 Seyferts 
(e.g. Boller, Brandt \& Fink 1996, hereafter BBF96)
and quasars (e.g. Laor et~al. 1994; Laor et~al. 1997). 
Ultrasoft narrow-line Seyfert 1 
galaxies (hereafter NLS1; 
see Osterbrock \& Pogge 1985, Goodrich 1989
and Stephens 1989) represent one 
extreme of this anticorrelation
and can often have \ROSAT photon indices, from simple power-law
fits, exceeding 3. Such steep spectra appear to be due to
strong soft X-ray excesses (compared to the underlying 
X-ray power law) in the \ROSAT band below
about 1 keV. These soft X-ray excesses show rapid and large
amplitude variability, supporting a compact origin.
NLS1 also tend to lie towards the
strong Fe~{\sc ii}, weak [O~{\sc iii}] extreme of the
first Boroson \& Green (1992) eigenvector, and
for this reason they are sometimes 
called I~Zw~1 objects.

Comparatively little is known about the harder X-ray
properties of NLS1, although \ASCA 
(Tanaka, Inoue \& Holt 1994) observations of a few
ultrasoft NLS1 have shown that they can have interesting
spectral differences from more typical Seyfert 1s. 
For example, Pounds, Done \& Osborne (1995, hereafter PDO95)
found a power law with a
remarkably large 2--10 keV \ASCA photon
index of $\Gamma\approx 2.6$ in the \ROSAT Wide Field
Camera NLS1 RE~J~1034+393. They also confirmed the 
presence of a strong soft X-ray excess above the
steep power law (also see Puchnarewicz et~al. 1995). 
This soft excess dominates the
X-ray spectrum at energies $\simlt 1$~keV. PDO95 drew an
analogy between RE~J~1034+393 and Galactic black
hole candidates accreting in their ultrasoft high states.
They also suggested that its intense soft X-ray flux may
be cooling its accretion disc corona to cause the
anomalously steep hard X-ray spectrum
(see section 7 of Maraschi \& Haardt 1997). It is 
important to determine if other ultrasoft NLS1 also
show steep 2--10 keV spectra 
(see Pounds \& Brandt 1997), as this could give 
much needed clues about the formation of the 
underlying continua of all Seyferts. 
 
In this letter we use the currently available \ASCA
data to compare the $\approx$ 2--10 keV 
intrinsic power-law slopes of NLS1 with soft \ROSAT 
spectra to those of Seyfert 1s with larger
H$\beta$ FWHM. We are interested in the {\it intrinsic\/}
slopes (free from effects such as Compton reflection) since 
these are thought to probe the fundamental nature of the 
underlying hard X-ray continuum. As is standard practice,
we quantify spectral slopes using power-law photon indices, and 
power laws usually appear to be
reasonable representations of the {\it intrinsic\/} 2--10~keV
continua, at least to first order.

\section{Spectral slope comparison} 

\begin{figure}
\centerline{\psfig{figure=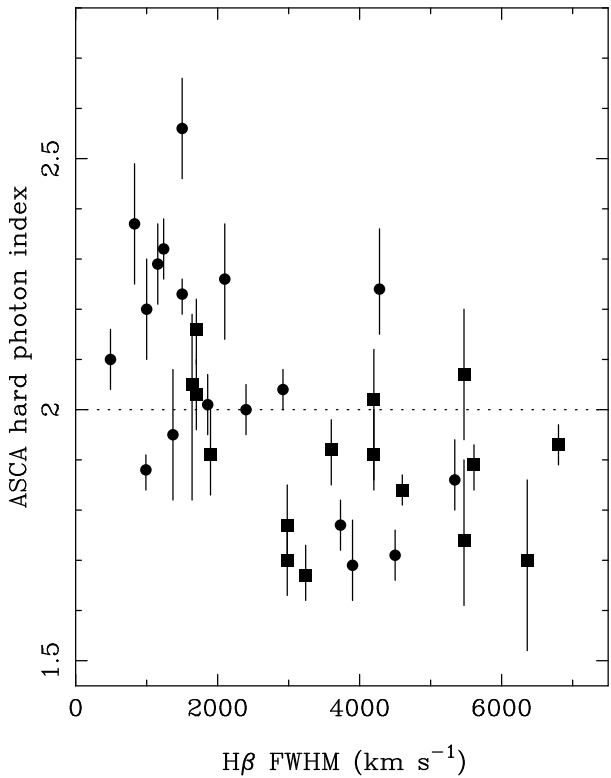,angle=0,width=0.5\textwidth}}
\caption{Plot of \ASCA hard X-ray photon index versus
full width at half-maximum intensity of the H$\beta$ line 
for Seyfert~1 type galaxies. Squares are Seyfert~1s 
from N97, and dots are other Seyfert~1s from the 
literature (see the text for details). The dotted line
at $\Gamma=2$ is drawn for illustrative purposes. 
}
\end{figure}

\subsection{Selection of objects under consideration}

Figure 1 shows a plot of \ASCA 
hard ($\approx$ 2--10 keV) photon index versus 
the FWHM of the H$\beta$ line. This plot may be 
compared with the analogous plots made using lower-energy
\ROSAT data shown as figure~8 of BBF96 and figure~5a
of Laor et~al. (1997). We 
have only used \ASCA X-ray data to make 
this diagram to avoid calibration uncertainties between 
different satellites. We have only used data above 2~keV
to avoid possible confusion by soft X-ray excesses,
cold absorbers, warm absorbers and other low-energy 
spectral complexity.

\subsubsection{The Nandra et~al. (1997) Seyfert 1s}

The squares in Figure 1 are Seyfert 1s from 
the large sample recently presented by 
Nandra et~al. (1997, hereafter N97). We have used 
the photon indices from table 6 of N97. These
are for the 3--10 keV band. The effects of Compton 
reflection have been included in the fitting, and this 
tends to increase the derived intrinsic photon indices  
by $\Delta\Gamma\approx 0.12$. Appropriate modelling 
of the iron K$\alpha$ emission line has also been 
performed, although the derived
photon indices are not sensitive to details of
the line model (see section~4.3 of N97). 
The 3--10~keV photon indices are expected to be very
similar to the 2--10~keV photon indices (K. Nandra,
private communication), and explicit spectral fitting of 
several of the N97 objects verifies that this appears to 
be the case. 

All objects in N97 are included in 
Figure 1 except for NGC~4151, NGC~6814
and 3C~120. We have excluded NGC~4151 due to 
its complex spectrum in the \ASCA band which hinders a 
precise determination of its intrinsic photon index
(see Weaver et~al. 1994 and references therein) as well
as its strong optical line profile 
changes (e.g. Penston \& P\'erez 1984). 
We have excluded NGC 6814 because it does not have  
measured photon indices in most of the tables of N97
(due to its faintness) and also 
because it shows persistent and strong optical 
line variability (e.g. Sekiguchi \& Menzies 1990).
We have excluded 3C~120 because it is radio loud and
has superluminal motion. We have not 
used the N97 results for NGC~4051 or Mrk~766, and 
we discuss these two objects in 
more detail below. Three objects in 
the N97 sample were observed by \ASCA twice
(MCG--6--30--15, Mrk~841 and NGC~3783), 
and we have used all six of 
these \ASCA measurements in our analysis. 
Our statistical results below do not materially change 
if we systematically use only the flattest, only the 
steepest, or the averaged N97 photon 
indices for these three objects. 
The N97 sample is composed of well-studied, hard X-ray 
bright Seyfert 1s, as these were the Seyfert 1s 
predominantly observed during early \ASCA observations. It 
is biased against soft \ROSAT NLS1 at some 
level. NGC~4051 and Mrk~335, the two objects in the N97 
sample with the smallest H$\beta$ FWHM, have their 
similarities to NLS1 discussed in section 5.1 of BBF96.  

We have verified that radio-loud objects,
which generally have flatter X-ray spectra than
radio-quiet objects, are not affecting the trend shown
in Figure 1. The N97 objects with large H$\beta$ FWHM 
are `typical' radio-quiet Seyfert 1 galaxies such as 
NGC~3227, NGC~3516, NGC~3783, NGC~5548 and Fairall~9.  

\subsubsection{Other Seyfert 1s and low-luminosity 
Seyfert~1 type quasars}

The solid dots in Figure 1, from left to right, show 
the 2--10~keV \ASCA photon indices for
IRAS~17020+4544 (Brandt et~al., in preparation),
PG~1244+026 (Fiore et~al., in preparation),
NGC~4051 (Guainazzi et~al. 1996),  
H~0707--495 (Hayashida 1997 and independent fitting),
NAB~0205+024 (Fiore et~al., in preparation),
I~Zw~1 (Hayashida 1997),
Mrk~478 (Madejski et~al., in preparation),
RE~J~1034+393 (PDO95),
PKS~0558--504 (Brandt et~al., in preparation),
PG~1211+143 (Yaqoob et~al. 1994 and independent fitting),
\Ir (Brinkmann et~al. 1997 and independent fitting),
Mrk~766 (Leighly et~al. 1996), 
PG~1116+215 (Nandra et~al. 1996 and independent fitting), 
ESO~141--G55 (Reynolds 1997 and independent fitting),
Mrk~1040 (Reynolds 1997 and independent fitting), 
RX~J~0437--470 (Wang et~al. 1997 and independent fitting),
MR~2251--178 (Reynolds 1997 and independent fitting) and
Mrk~290 (table 2 of Turner et~al. 1997). 
Iron line modelling has been included in the fitting 
when appropriate (see the cited references for details).
As with the N97 Seyfert~1s, the photon indices 
for these objects do not appear to be sensitive to 
the details of the iron line modelling.

The intrinsic hard photon indices of NGC~4051 and Mrk~766 
appear to be variable. For NGC~4051, we have used the
photon index from the time averaged analysis in section 4.3 
of Guainazzi et~al. (1996). This photon index appears to 
provide a reasonable description of the typical source
behaviour (see figure 12 of Guainazzi et~al. 1997). 
For Mrk~766, Leighly et~al. (1996) 
state that the 2--10~keV photon 
index varies abruptly between $1.57^{+0.07}_{-0.07}$ and 
$2.00^{+0.03}_{-0.04}$ when the source changes from a
low state to a high state. The time averaged value is 
likely to be closer to 2.00 than 1.57 
(see Leighly et~al. 1996 for details), and table 2 of 
N97 gives a time averaged value of $2.00^{+0.05}_{-0.05}$
(a similar value is obtained by Hayashida 1997). 
We adopt this value and show below that our statistical 
results are not sensitive to the precise photon index 
adopted for Mrk~766. All results discussed 
below should be implicitly taken to
pertain to time averaged Seyfert 1 behaviour,
although most of the Seyfert~1s in our sample did
not show strong photon index variability during their
\ASCA observations. 

Figure~1 does not include the radio-loud objects from 
Reynolds (1997) or highly luminous quasars
(with 2--10~keV luminosities larger than 
$1\times 10^{45}$ erg s$^{-1}$) as our interest here is 
in radio-quiet Seyfert~1 galaxies. 
We have not used NGC~2992 (a narrow emission line 
galaxy with internal absorption; NELG)
from the Reynolds (1997) sample, due to the fact 
that its intrinsic photon index is not well determined
by the \ASCA data (see Weaver et~al. 1996; also see
Section~1 for an explanation of why we are only 
interested in intrinsic photon indices). The 
2--10~keV spectrum of NGC~2992 is strongly influenced by 
a time-delayed Compton reflection continuum and other 
complexities which obscure the true shape of the
underlying continuum.
We have not plotted the NLS1 IRAS~13224--3809 
(see Boller et~al. 1993) in Figure 1 due 
to the fact that its 2--10 keV spectral shape appears to 
be highly variable and is poorly constrained by current 
data (see table 4.4.1 of Otani 1995). Its hard X-ray photon 
index appears to vary between roughly 1.3--2.3, although the
precise values obtained depend on  
details of the modelling. For comparison with
Figure 1, IRAS~13224--3809 has an
H$\beta$ FWHM of $\approx 650$ km~s$^{-1}$, and   
Hayashida (1997) gives a time averaged hard 
photon index of $2.15\pm 0.13$.

\subsubsection{Selection effects}

While the objects in Figure 1 are not drawn from a 
rigorously defined complete sample, we have used all
the currently available radio-quiet Seyfert 1s with 
reliably measured {\it intrinsic\/} \ASCA photon indices 
in an attempt to avoid any biases. There are 34 data
points in Figure~1. 

One possible selection effect is that several
of the NLS1 that have been observed with \ASCA 
have been chosen as targets due to their 
soft spectra in the \ROSAT band. This does not 
invalidate comparisons made using the 
data in Figure 1, but it does restrict
our comparison to that of soft \ROSAT spectra NLS1 
versus Seyfert 1 galaxies with larger H$\beta$ FWHM. 
This is to be understood in all discussion
below. If the \ROSAT $\Gamma$ versus H$\beta$ FWHM
correlation, after correction for possible 
absorption effects in the \ROSAT band, is as tight 
as suggested by figure 5a of Laor et~al. (1997), 
then a distinction between soft \ROSAT NLS1 and
NLS1 more generally may be unimportant.

Seyfert 1 galaxies with both \ROSAT photon indices 
larger than 3 as well as H$\beta$ FWHM larger than 
3000 km s$^{-1}$ appear to be strongly discriminated
against by nature (see Section 1; Pounds \& Brandt 1997
give the clearest illustration of 
this effect so far). Thus the lack 
of \ASCA observations of Seyfert 1s with both 
large H$\beta$ FWHM and \ROSAT photon indices 
larger than 3 is not an artifical 
selection bias, but is rather due to a 
real and physically interesting effect. 

\subsection{Statistical comparisons} 

Examination of Figure~1 suggests that NLS1 with soft 
\ROSAT spectra have $\Gamma_{\rm Hard}>2$ more 
frequently than Seyfert~1s with larger H$\beta$ FWHM. 
Below we perform statistical tests to quantify the 
significance of the apparent trend in Figure~1. 

\subsubsection{Definition of samples} 

Some of the statistical tests below are designed to
compare two samples of objects. Therefore, we
have divided the Seyfert 1s in Figure 1 into
two samples. The first sample (hereafter NL
for `narrow lines') is composed of the 
objects with H$\beta$ FWHM 
smaller than 2000 km s$^{-1}$. The second sample
(hereafter BL for `broad lines') is 
composed of the objects with  
H$\beta$ FWHM larger than 2000 km s$^{-1}$.
We have chosen the 2000 km s$^{-1}$ dividing value of
H$\beta$ FWHM based on the a priori definition 
for NLS1 given in section IVb of Stephens (1989).
However, we have included \Ir in NL rather than BL.  
As discussed in section 4.5 of 
Brandt, Fabian \& Pounds (1996), which was
written before the \ASCA results were
available, \Ir has many striking similarities
to NLS1. In addition to its rather narrow Balmer
lines (it has an H$\beta$ FWHM of 2100 km s$^{-1}$), 
\Ir has a soft \ROSAT spectrum, strong optical Fe {\sc ii} 
emission and weak forbidden line emission
(see Wills et~al. 1992). As our
goal below is to compare the \ASCA hard photon index 
distributions of soft \ROSAT NLS1 and Seyfert~1s
with larger H$\beta$ FWHM, inclusion of \Ir in 
NL is clearly warranted. We show later that
our conclusions do not depend on the treatment
of \Irc. In total, we have 15 photon indices in NL 
and 19 photon indices in BL. 

\subsubsection{Statistical tests}

We first performed a two-sample Kolmogorov-Smirnov test 
to determine the probability that NL and BL 
were drawn from the same parent distribution 
function. This test is robust and nonparametric, 
although it does not provide formal information 
as to why two distributions are different 
(i.e. they could differ in mean, variance,  
skewness or some other property, but the 
Kolmogorov-Smirnov test does not discriminate
between these possibilities). 
NL and BL are large enough for the
Kolmogorov-Smirnov test to be applicable (see equation 
14.3.10 of Press et~al. 1992 and Stephens 1970).
We obtain a Kolmogorov-Smirnov statistic of $D=0.604$, and
in Figure~2 we plot the cumulative distribution 
functions for NL and BL used by the
Kolmogorov-Smirnov test. Given the value of $D$ we 
find that NL and BL have less than a 0.25 per cent 
chance of being drawn from the same parent distribution 
function, and Figure 2 suggests that the main
reason for the low probability is likely to be a 
difference in the central photon 
indices of NL and BL. 

\begin{figure}
\centerline{\psfig{figure=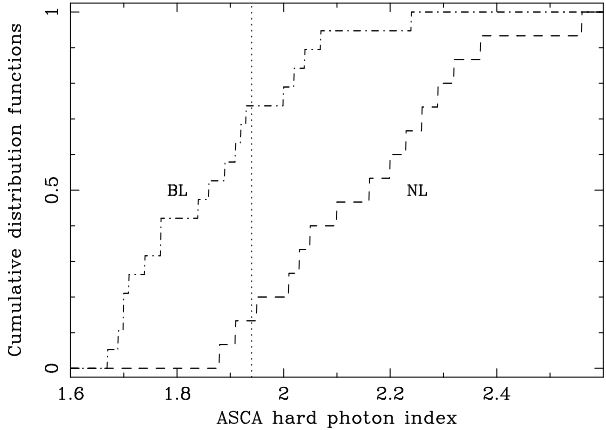,angle=270,width=0.5\textwidth}}
\caption{Cumulative distribution functions for the 
photon indices of NL and BL. The vertical dotted line
shows the value of the photon index
at which the Kolmogorov-Smirnov $D$ statistic is obtained. 
}
\end{figure}

In order to obtain more descriptive information 
about the differences between NL and BL, we have
calculated some of their moments.
The mean photon index of NL is 2.15,
the variance of the NL photon indices is 0.036 
and the standard error of the NL mean is 0.049.
The mean photon index of BL is 1.87,
the variance of the BL photon indices is 0.025 
and the standard error of the BL mean is 0.036.
Note that the NL mean photon index is $\simgt 5$ 
standard errors larger than the BL mean photon
index. 

We have used Student's t-test (for distributions
with nonequal variances; see Press et~al. 1992) to
examine whether NL and BL have significantly 
different means. We obtain a t value of 4.73. 
There is less than a 0.02 per cent chance that
t could be this large or larger, for distributions
with equal means. While Student's t-test is 
a parametric test and is 
subject to the limitations described in chapter 3
of Siegel (1956), we nevertheless consider the
large value of t to be suggestive.

We have performed a Spearman rank-order 
correlation using the data shown in Figure 1. 
This test is sensitive to any monotonic relation
between two variables and is more general than
Pearson's r-test. As described in detail in 
Siegel (1956), it is nonparametric, robust and
requires far fewer assumptions than parametric tests. 
It also does not require the division of the objects
into NL and BL (this division is somewhat artificial
if there is a continuous trend present in Figure~1).  
We have checked our implementation of the Spearman test
using data from Laor et~al. (1997), and
we obtain entirely consistent results. Using the data
in Figure~1, we obtain a Spearman $r_{\rm s}$
value of $-0.583$. The Spearman probability 
associated with this $r_{\rm s}$ value 
is $3\times 10^{-4}$ (that is, there 
is less than a 0.1 per
cent chance of obtaining a Spearman $r_{\rm s}$
value with this large a magnitude by chance).
Thus, the Spearman rank-order
correlation indicates that there is a highly
significant anticorrelation between H$\beta$ FWHM
and \ASCA hard photon index for the objects in 
Figure 1.
 
\subsubsection{Safety checks}

We have performed several safety checks to examine the
reliability of our statistical results. First of all, we
have verified that they do not depend on any single 
data point in Figure~1. This is understandable since the
Kolmogorov-Smirnov and Spearman tests we have employed
are relatively insensitive to outlying points. For
example, we still find strong evidence for an anticorrelation 
even if the most extreme NLS1 illustrated, RE~J~1034+393, 
is neglected. Running the Spearman rank-order 
correlation without RE~J~1034+393, we 
obtain an $r_{\rm s}$ value of $-0.566$ and a Spearman
probability of $6\times 10^{-4}$. However,  
there is no known reason why RE~J~1034+393, or any other
object in Figure~1, should be neglected. 

Conservatively using the flat photon index for
Mrk~766 from section 4.1 of Leighly et~al. (1996)
does not change our basic statistical results. Using
the flat photon index, we obtain $r_{\rm s}=-0.549$ and
a Spearman probability of $8\times 10^{-4}$.
We have also verified that the inclusion of 
NGC~2992 or IRAS~13224--3809 (see Section 2.1.2)
does not change our basic statistical conclusions. 
Including \Ir in BL rather than NL does not affect the 
anticorrelation, because the Spearman analysis does not depend
on the division into NL and BL.

If we use the hard photon indices 
from table 4 of N97, which do not include Compton 
reflection in the spectral fitting, the trend suggested 
above is not affected. In fact, it becomes statistically 
stronger because all the squares in Figure~1 shift 
downward (on the average by $\Delta\Gamma=-0.12$). 

As the photon indices in Figure 1 have errors 
associated with them, we have performed Monte-Carlo 
simulations to verify that it is not just a 
chance realization of the best-fitting 
photon index values that leads to the apparent 
anticorrelation. In these simulations we generate
sets of photon indices based on the data points
and errors in Figure 1, and we then perform Spearman
rank-order correlations using these sets. 
We generate the photon indices from Gaussian 
distributions centred on the points in Figure 1, and 
we follow the conservative prescriptions in 
section 2.1 of N97 regarding the translation of 
confidence limits into Gaussian $1\sigma$ 
values. 
In 94.0 per cent of the simulations, we
can rule out a chance anticorrelation with
greater than 99 per cent confidence. 
In 99.7 per cent of the simulations, we
can rule out a chance anticorrelation with
greater than 95 per cent confidence. 
Note that the 95 per cent confidence level
stated here does not mean that we have weakened our 
threshold for correlation testing. It is 
only used to describe the (in itself  
unlikely) tail of the distribution of 
Spearman probabilities that arises from our 
Monte Carlo calculations.
It is unlikely that our statistical results 
are due to a chance realization of the best fitting
photon index values. 

\section{Discussion} 

The currently available data suggest, with high
statistical significance, that soft \ROSAT 
NLS1 have generally steeper 2--10 keV
intrinsic spectral slopes than Seyfert 1s with larger 
H$\beta$ FWHM. Note that this is a robust statistical 
statement regarding two populations, and that the discovery 
of a single discrepant Seyfert~1 is unlikely to 
alter it. 

The trend suggested by Figure~1 appears similar to the
well-established \ROSAT $\Gamma$/H$\beta$ FWHM trend,
although the range of photon indices in Figure~1 
($\approx$ 1.6--2.6) is about half that
seen in lower-energy \ROSAT data 
($\approx$ 1.9--4.2). Soft 
excesses probably lead to the larger photon index
dispersion observed in the \ROSAT band. However,
soft excesses are thought to be confined
to below about 1~keV. Figure~1 uses no data 
below 2~keV, and thus the hard photon index trend 
illustrated there would probe different and new
Seyfert physics. As discussed by PDO95, photons 
from the soft excesses of ultrasoft NLS1 could
Compton cool the coronae which create their harder
flux and thereby steepen their 
2--10~keV continua (also see Maraschi \& Haardt 1997).
If this is indeed the cause of the 
observed effect, then comparative 
studies of soft \ROSAT NLS1 and Seyfert~1s with 
broader permitted lines will have provided one of the 
first direct illustrations of a physical process that 
influences the origin of the underlying hard X-ray 
continua from Seyfert galaxies. Other
possibilities, such as significant curvature in
the underlying 2--10~keV continua of some Seyfert~1s,
also need further examination (for a detailed study
see Brandt et~al., in preparation).

Some researchers have considered models for NLS1 
in which they are Seyfert 1s viewed along a particular 
line of sight (e.g. highly pole-on; Osterbrock \& Pogge 1985; 
Puchnarewicz et~al. 1992), although such models appear to 
have trouble explaining all the known optical line 
(e.g. Boroson 1992; but also see Hes, Barthel \& Fosbury 1993 
for evidence that [O~{\sc iii}] may not be isotropic)
and X-ray properties.  
If, despite current indications, a special orientation 
does turn out to be the root cause of NLS1 properties 
then the hard photon index trend described here would
require the hard power-law shapes of Seyfert 1s to
be orientation dependent. 

Additional \ASCA analyses of NLS1 
(e.g. Ton~S~180, RX~J~0148--27 and IRAS~20181--2244) 
are needed to further study the hard photon index 
trend suggested here. They will thereby probe the full 
range of Seyfert 1 spectral shapes (cf. Elvis 1992) and 
clarify the extent to which historical and other selection 
effects have influenced measurements of the dispersion of 
Seyfert 1 hard X-ray slopes. It would also be of
great interest to see if an analogous trend is found
in higher luminosity Seyfert 1 type quasars. The
Laor et~al. (1997) sample is an obvious one in which
to search for such a trend, although it may not have 
enough ultrasoft quasars to provide definitive results.
There are only three quasars in Laor et~al. (1997)
with best fit \ROSAT photon indices larger than 3, and one
of these is faint and unlikely to yield precise
\ASCA spectral parameters.

\section*{Acknowledgments}

We gratefully acknowledge financial support from the 
Smithsonian Institution (WNB) and 
NASA grant NAGW-4490 (SM).
We thank
F. Fiore 
for communicating two photon indices 
prior to publication.
We thank
C. Done,
A. Fabian,
F. Fiore,
J. Halpern, 
A. Laor,
K. Nandra,
R. Pogge, 
C. Reynolds and 
B. Wilkes 
for helpful discussions.

\bsp

\end{document}